# Higher School of Economics
# (National Research University)

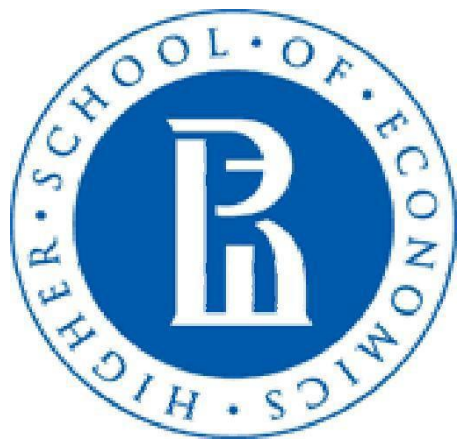

**Faculty of Computer Science**

**Department of Software Engineering**

**Master Thesis**

**Transparent Voting Platform Based on Permissioned Blockchain**

Supervisor:

Associate professor

Nikolay Kazantsev

Academic Consultant:

Ilya Eriklintsev

Presented by:

Nazim Faour

# Contents

# ABSTRACT

Since 2004, different research was handling the challenges in the centralized voting systems, e-voting protocols and recently the decentralized voting. So electronic voting puts forward some difficulties regarding the voter anonymity, the secure casting of the votes and to prevent the voting process from frauding. The Decentralized property of the technology called "blockchain" could have the solution for many of the challenges in voting research area and brings a new secure mechanism of safe and transparent voting. In this paper, a broad comparison between ongoing voting systems has studied by analyzing their structure and the drawbacks that should consider in future to improve the whole election process from keeping the privacy of the voter, casting a vote with the possibility to check if it was counted correctly to publishing the results. The result of the paper will give a new approach to extend the target of the election from small scale to large scale despite the fact of Ethereum limitation which can cast on the blockchain just five votes per minute. The primary challenge is to find an answer for this question: "How to balance between voter privacy and transparency without breaking the important rule where the voter can proof for a specific candidate that he voted for him in a bribe situation?".

# Chapter 1

# Introduction and Literature Review

## 1.1 Introduction

The protection of integrity of digital part of information requires the blockchain technology which is a decentralized and distributed database in a peer to peer network. In blockchain system the data is shared between all the nodes of the p2p network. The data is stored with considering the maximum size and the verification by using a specific technique for hashing. This hashing technique will contain a specific number of zeros at the beginning which represent how many participants does the system has in the network. Transactions are the real data in a blockchain system which are totally public. If the user tries to make a transaction (sending, receiving bitcoins or casting a vote), the system will verify the transaction before adding it to the blockchain. So this verification will prevent the double spending or the fault votes.

| Ledgers of intermediaries e.g. Bank account | Blockchain e.g. Bitcoin |
|---|---|
| Centralised: Single owner such as a bank makes these ledgers vulnerable since they have a *single point of failure* that can be hacked. | Distributed: Nobody/everybody owns it. Because it is distributed across millions of users it has *no single point of failure* making it especially difficult and economically unfeasible to hack. |
| Opaque: Only authorised users can view them or have access to them. | Transparent: Anyone can view or access the entire ledger, which is updated in near-real time. |
| Alterable: Errors can be corrected by (internal) users with overriding privileges. | Immutable: Transactions cannot be reversed. |
| Subject to identity theft: Accounts are often hacked. | Encryption and pseudo-anonymity. This makes it very difficult to hack blockchain. |
| Time lag: Can take days or even weeks to complete transactions. | Near-real time: Transactions completed in ten minutes (on average). |
| Borders: Varying international and conversion fees that can cost up to 20 per cent. | Borderless: Same low fee everywhere (usually a few US cents). |

Table 1.1 comparing blockchain vs traditional ledger

In another words, the blockchain can be defined as a list or a decentralized ledger of all transactions that are procced in a p2p network. Blockchain technology is used in Bitcoin and the other current cryptocurrencies.

| Category | Question | Bitcoin's approach | Other ways |
|---|---|---|---|
| Data storage | How should data be stored? | A blockchain | A database (could be replicated across multiple data centres) |
| Data distribution | How should new data be distributed? | Peer-to-peer | Client-server, hierarchical |
| Consensus mechanism | How should conflicts be resolved? | Longest chain rule | (Not needed in trusted networks) 'Trusted' or super-nodes |
| Upgrade mechanism | How do the rules get changed? | BIPs (for writing the rules) Vote by hashing power (for implementing the rules) | Centralised upgrades Contractual obligations |
| Participation criteria | Who can submit transactions? | Pseudonymous, open | Trusted, pre-vetted participants |
| Participation criteria | Who can read data? | Pseudonymous, open | Trusted, pre-vetted participants |
| Participation criteria | Who can validate transactions? | Pseudonymous, open | Trusted, pre-vetted participants |
| Participation criteria | Who can add blocks? | Pseudonymous, open | Trusted, pre-vetted participants |
| Defence mechanism | How to prevent bad behaviour? | Proof-of-work | (Not needed in trusted networks) Proof-of-stake, other 'proofs' or costs to add blocks |
| Incentivisation scheme | How to incentivise block-makers? | (only expensive in Bitcoin because of proof of work) Block reward, to be replaced by transaction fees | Contractual obligations 3rd party funding |
| Incentivisation scheme | How to incentivise blockchain data storage? | Not considered | Contractual obligations 3rd party funding |
| Incentivisation scheme | How to incentivise transaction validators? | Not considered | Contractual obligations 3rd party funding |

Table 2.2 bitcoin and blockchain

In any election, Threats are always exist even if the process of election is paper traditional one or electronic one (e-voting) due to the importance of the results of an election and the high level of stakes for the one who will win the election.

On the last decade, a lot of election results has been fraud. The fraud includes some attacks such as double voting, buying the vote and using the blank ballots. So the question is," how to be sure about the results of the election that it's correct and how to find out if it's wrong?".

In paper voting, there is always a trusted party which is responsible of counting the votes and the voters must rely on that. in this type of elections ,the whole process of verifiability and tallying performed only by the trusted party so the voters cannot find a way to check and verify the correctness of the final results.

In "end to end voting verifiable systems", this whole dependency on a trusted party is reduced in order to give the right to the voter to check and verify the results if it's correct or not.

## 1.2 Literature Review

"Permissioned Blockchain" means that nodes must have former permission from a centric authority in order to make any changes to the ledger.
Using Blockchain as a distributed database for p2p voting system will give transparency due to a reason that the network of nodes will be public and it can take a huge amount of the total computing power in order to modify or change some piece of information which is stored on the blockchain. In Addition, this technology will allow the data to be transparent and not susceptible to corruption.  The fact about that blockchain does not have a failure of single point, will make it most suitable for a voting system. This system will be able to verify the quality for each vote to be totally authentic so any election will be secure and transparent.

The blockchain can give an exceptionally large and scalable solution to the current voting methods with increasing the security and fraud-proof digital voting.

There are many advantages for using a blockchain, which make the blockchain a secure replacement to the other databases.

- High Availability: many nodes totally distributed and storing the whole database.
- Integrity and Verifiability: each chain is verified and then attached to the blockchain. So any altering to some block will effect the whole chain and every block should be recalculated which sound impossible.
- Easy to define one common starting point, where to store the data, always attached it to the last block in the longest chain.

All previous advantages lead to build a voting system with blockchain technology.

| N | Keyword | Hits | Selected |
|---|---|---|---|
| 1 | Decentralized Database | 2640 | 2017 |
| 2 | Blockchain Technology | 1170 | 2017 |
| 3 | Blockchain bitcoin | 580 | 2017 |
| 4 | End to end verifiable voting systems | 3510 | 2016-2017 |
| 5 | E-voting | 1310 | 2016-2017 |
| 6 | Voting with blockchain | 418 | 2016-2017 |

Table 1.3 Sources for the research idea

| N | Name of authors | Research area | Statement |
|---|---|---|---|
| 1 | Dr. Jeremy Clark | End-to-End Verifiable Voting Systems. Bitcoin and Blockchain | "Requiring users to manage cryptographic keys has been shown through usability experiments to be difficult" |
| 2 | Dr. Jeremy Clark | End-to-End Verifiable Voting Systems. Bitcoin and Blockchain. | "If voters generate or are provided cryptographic keys to use in the voting process, hackers will concentrate on compromising these keys through interception or malware." |
| 3 | Dr. Jeremy Clark | End-to-End Verifiable Voting Systems. Bitcoin and Blockchain. | "A [sic] voting system that uses a blockchain as a public ledger but requires voters to show up and vote in person is an excellent option for elections today, but reaching beyond that is too risky." |
| 4 | Dr. Feng Hao | co-lead of the Secure & Resilient Systems group at Newcastle University's School of Computing Science | "end to end verifiable voting systems have the merit of allowing a voter to verify if their vote is correctly recorded and correctly included into the tallying process—and if ballots are missing in transit or modified, it can be detected by voters." |
| 5 | Patrick McCorry | Voting with blockchain | "Everyone can cast their encrypted vote. And then at the end of the election, once all the votes have been cast, anyone, including observers, can simply add the encrypted votes together. It will cancel out all the random factors in the encryption and it will just reveal the final tally." |

Table 1.4 Review of authors

Creating a voting system over blockchain provides a platform which is entirely trusted to be able to have data verification process in real time and to have automated execution of specific voting protocols.

## 1.3 Challenges and Risks:

- Each technology has its difficulties and the same logic implements on blockchain, it's normal to have some difficulties at the beginning of the blockchain evaluation, for example, the process of verification and the exact speed of each transaction.
- Different issues in the field of cyber security still exist and those issues should be solved in order to bring the blockchain technology to the trust in the real world where everyone will trust the system to put their data into it.
- The action and process of integrating concerns, Blockchain applications offer many solutions that need big and important changes or complete changing of the current systems. So To make this changing step, companies must devise a strategy for the transition.

- Adoption for the new system, in order to implement a blockchain concept the whole existing system must be transferred into decentralized p2p network.
- Blockchain brings a huge savings in the time cost and transactions cost but on the other side the initial step can require a high costs.

# Chapter 2
# State of the Art and Current Research

An important challenge in cybersecurity field was hold by the Economist and Kaspersky Lab in September 2016 and 20 worldwide universities participated in this challenge. The challenge was on how to have a secure digital voting with the use of blockchain technology [20].

The first voting system which based on cryptographic and mix protocol was proposed by Chaum [26]. As a centralized voting in remotely condition, Some systems exist such as Civitas [27], DRE-i [28], Adder [29] and Helios [30]. Another voting systems in the condition of polling station are MarkPledge [31], Prêt `a Voter [32], Votegrity [33], DRE-ip [34], STAR-vote [35] and Scantegrity [36]. For decentralized voting systems, Groth [37] and Kiayias-Yung [38]. The only systems without tallying authority are DRE-ip and DRE-i.

In this chapter, different research in the area of voting will be introduced.

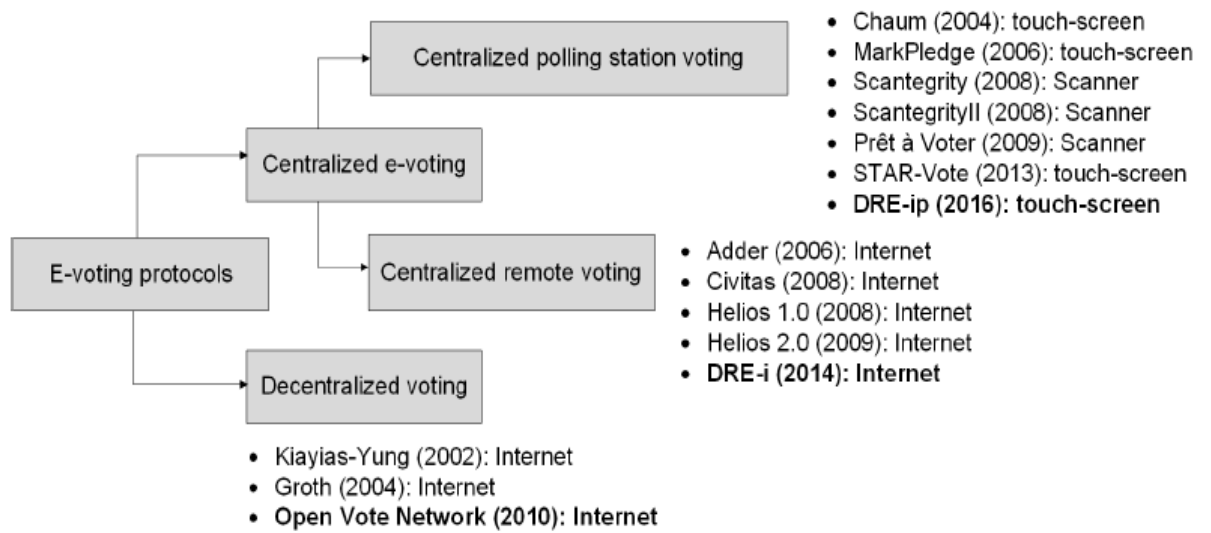

Figure 2.1 e-voting systems.

## 2.1 Votebook (New York University):

The votebook solution won the competition with the use of blockchain technology as a "permission blockchain" and without the use of the mechanism proof of work (PoW). Their proposal allows a centralized authority to be responsible for the way that the encryption keys are distributed on the network nodes and due to this reason was the need to use a permission blockchain. Each node in the network is a voting machine. Every voting machine will generate "public keys" and "private keys". "Private Key" will be stored under a secure matter, and the "public key" will be sent to a centralized authority.

The block that is proposed to be in the network will contain three parts:

1) Unique identifier for the node
2) Time stamp by using a time-based protocol
3) Validation process: A hash of the previous block, a set of voters with their vote and a digital signature.

Votbook has different considerations in the design process of the voting system:

1. The ability for each voter to check if his/her vote has been counted in a correct way.
2. No possibility for coerce in the electronic voting system.
3. The voting system should be able to handle the publishing of results or the hiding of rounds results as required.
4. The voting system must deal with the empty votes and not make those abstinence votes to be used in the counting process.
5. The voting system must be audible.

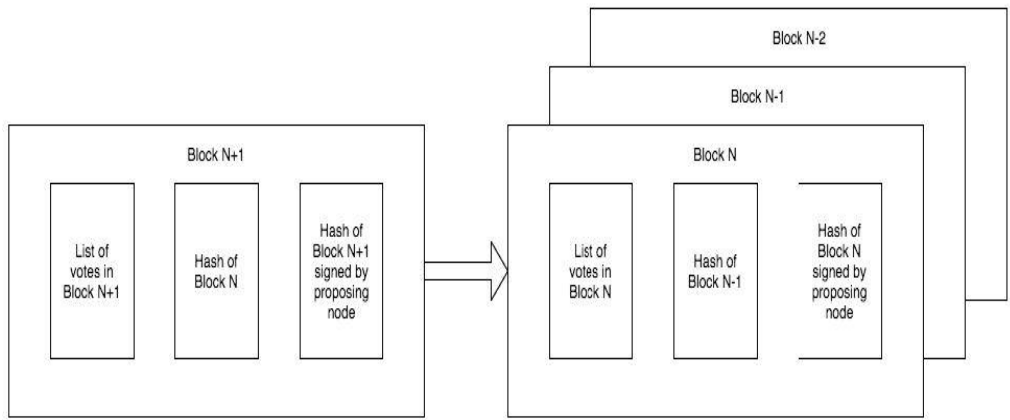

Figure 5.1 Block Structure in the network.

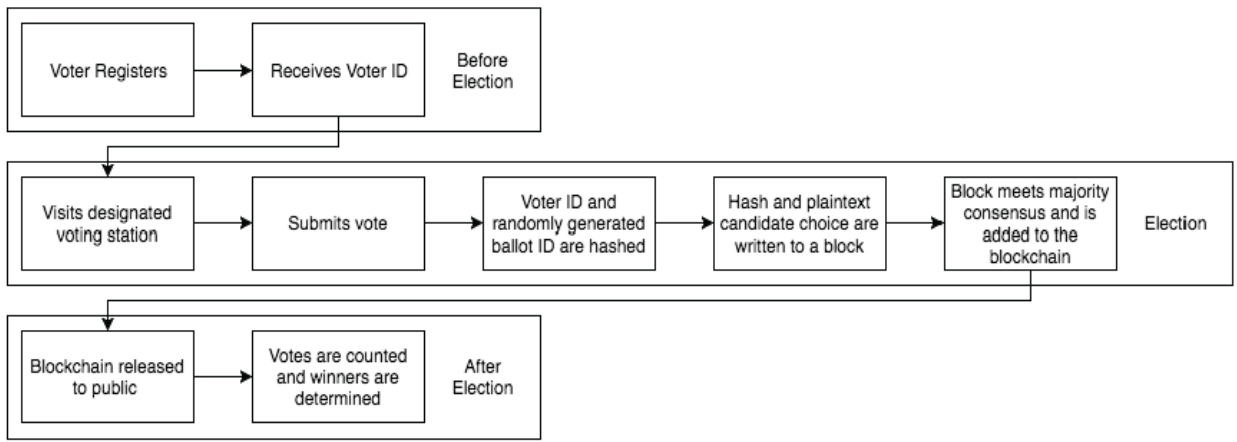

Figure 2.2 the Process of voting in Votebook.

Some Challenges that should be solved in the Votebook proposal:

1) The system does not solve threats that may face electronic voting.
2) Individual voting machines can still be tampered with or just denied service.

3) Not clear enough which hashing algorithm will the system use to hash the voter Id and the Ballot Id, how the private and public keys will be generated.
4) How the system will face the Sybil Attacks.
5) How the system will verify identity without sacrificing anonymity.

## 2.2 Open Vote Network (New Castle University):

The Newcastle University team proposes a decentralized voting system where the trusted authorities are removed from the process of the election. The proposal focused on the possibility to have electronic voting protocols with the use of Ethereum blockchain as a self-enforcing system.

The votes are cast on the distributed peer to peer network in multi rounds, and the voters verify the last tally but in a private way without getting any information about the other votes. This scenario is suitable just in the elections as small scale due to the fact of multi interaction rounds.

OV-net has many properties [22] :

1) Decentralized with a voting scheme of two rounds [23].
2) The tallying process gave the privilege to each voter to tall votes which called "self-tallying."
3) Implementing a proof of concept solution to work with Ethereum blockchain.
4) Two smart contracts: one is voting contract and the other cryptography contract.
5) Three html pages: election administrator, voter and observer.
6) Five stages of elections: setup, signup, commit, vote and tally.

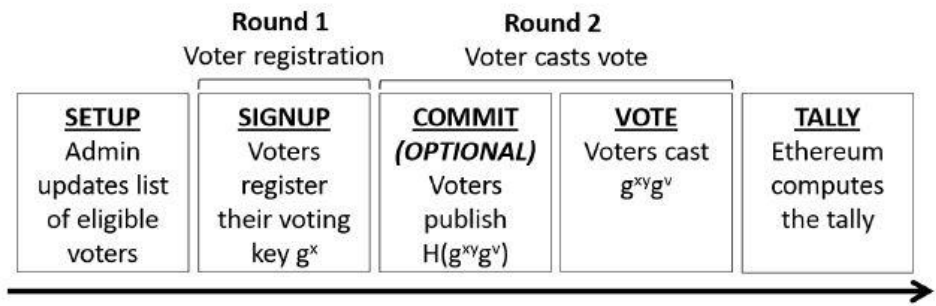

Figure 2.3 voting Rounds

In the setup stage, the admin is responsible for checking the authentication of voters with putting the whitelist of eligible voters and then decide the timing for the next steps with attaching the details such as the voting question and the registration fees.

The voters read the voting question and then decide to register by deposit on Ethereum in the signing up step.

There is an optional step which is commit step to ensure that the voters are commitment to their choice by sending their hash of the data of second round on Ethereum.

The next step is casting a vote and when the final vote is casted the admin notifies the blockchain Ethereum to calculate the tally.

Finally the results are published on the blockchain in the last step.

One of the Challenges in the OV-net was By Including an Elliptic Curve cryptography library, performing the process of computation becoming massive to store it on the Ethereum Blockchain due to the reason that solidity language does not support the Elliptic Curve cryptography.

## 2.3 The proposal of the University of Maryland:

Their proposal used the Ethereum blockchain to record the votes with the use of ZKP and Merkle tree as cryptographic primitives.

The Merkle tree proves to the voter that his vote included in the counting process after the end of the elections.

The ZKP proves the correctness of the tally process.

In their system, each voting machine represents a voter with a server that is responsible for handling decryption and tallying process.

The voter client here encrypts the vote with the "public key" of a centralized authority and then this authority is handling the decryption and tallying process in a correct and verifiable way. The proposal did not use a cryptographically approach to the cast of the vote but in fact they used a random number as a receipt.

They implement their idea by using Hawk to run the smart contract, manager and user code. The smart contract in their proposal will be tall every vote with spending coins in the voting process while choosing the candidate [24].

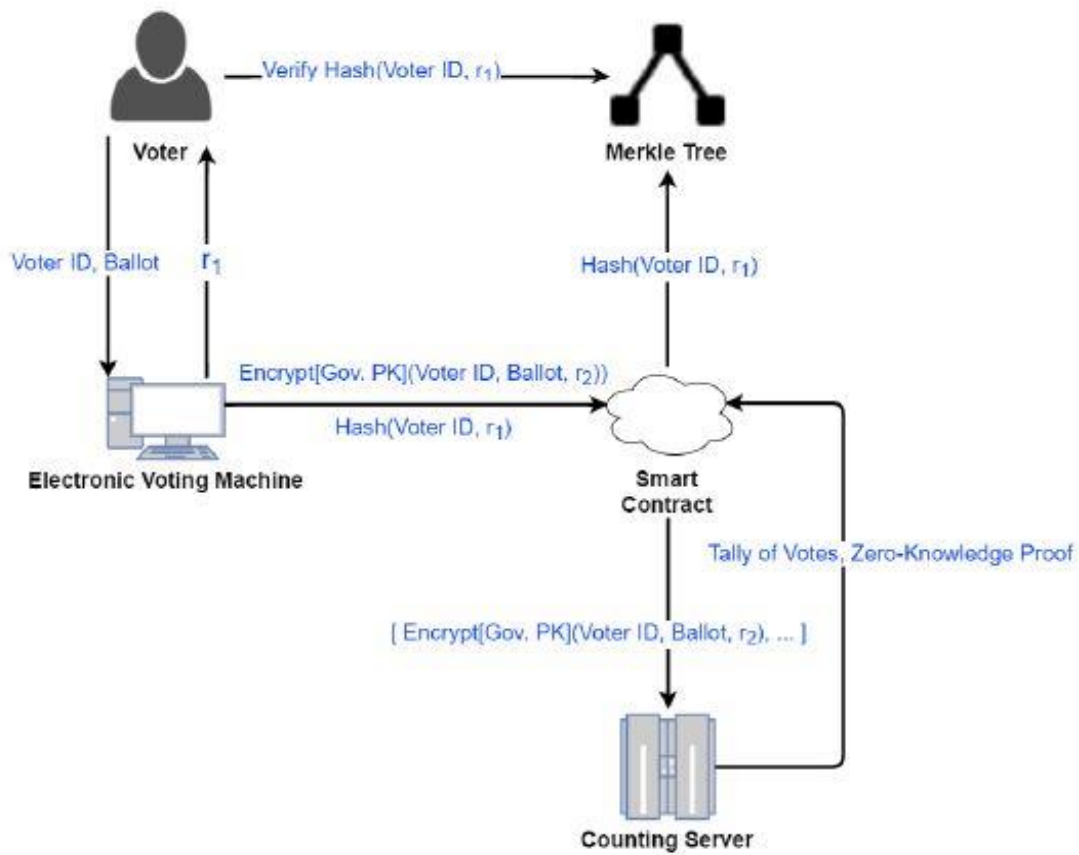

Figure 2.4 the voting proposal.

There is no guarantee if the vote cast correctly or was a part of the process of tallying even if the voter found his voteID in the blockchain. Also, there is no possibility to have a checking way to find if the vote was cast as wanted to be. Each voter will not have the right to find if his vote cast to his choice of candidate due to the reason of the encryption of candidate choices with the use of DRE.

## 2.4 The Voting under Unconditional Integrity and Privacy Concordia University:

They introduce a system with different properties of security and give a new vision how there are various interdependent combinations of security issues. Their system is depending on Eperio, which will let the voters cast ballots in paper-based and then leave without thinking or involving in the process of tallying [25].

One of the drawbacks is removing the possibility for the voter to be involved in the process of tallying so the voters must trust the honesty of the shareholders.

# Chapter 3

# End to End Voting Systems

In 1981, David Chaum presented his research idea about secure voting by using public key cryptography. He gave a technique which depend on the cryptography public key to make the participants identity unknown and hidden for the public communication. This technique it's not secure enough to be implemented in real world elections [2]. Chaum technique became used in different research areas but it did not have a chance to be tested in real world projects related to voting and elections [3].

The life cycle of the vote can be described in the following steps in order for the voter to be sure that if any breakdown or tampering happened in the system the voter will be able to discover that in those steps:

## 3.1 What should be included to keep voter privacy

- The secret issues for a ballot, this method provides anonymity for the voter choice in order to protect his privacy. So the system must not give any information related to the choice of the voter in casting stage.
- Receipt Freeness, it's about how the voting system can avoid giving details to the voter where he/she can use it in bribery way to a third party to prove that he/she voted as needed [7] [8].
- Coercion Resistance, this definition explains how to protect the voter and give him a secure environment to cast his vote even if there is a dealing with a specific coercer [1].

So this steps can be summarized as follows, The voter confirms that his vote has been encrypted correctly by the system. The voter confirms that his vote has been recorded correctly by tracking it with his/her receipt. The voting system will publish the cryptographic proofs of the correctness of the operation to ensure results integrity [1].

## 3.2 The verifiability in the end to end voting systems has three main steps

- Cast the vote as planned, in this step the one who voted can have the right to verify that his/her choice of the candidate on the ballot was correctly marked in the voting system.
- Record the vote as casted, in this step the voter can check if the voting system has recorded his/her vote correctly.
- The vote Tallied as it was recorded, In this step the voter can check if the voting system count his/her vote as recorded.

# Chapter 4

# General characteristics of a voting system

In a good voting system, these characteristics must be considered:

## 4.1 Integrity:

The condition of the whole voting system to be unified should be always guaranteed [4]. So the system ensures that no vote was changed under any case in the whole election process. No trust will be given to the system if it does not have integrity.

## 4.2 Eligibility:

In the voting process just the voters who are eligible can cast a vote .

Each voter can cast his vote once and no possibility for multiple times of voting.

## 4.3 Availability:

One of main properties of a voting system is the ability of this system to remain available in real time while the process of elections is going on. The voters should have the ability to check the results by using their physical devices.

The system should be able to handle large workload because some voters will cast their votes in simulate way.

## 4.4 Fairness:

Authority and fairness is an important specification of a voting system because the system should not publish any partial results before the time of voting ended in order not to give the voter the chance to modify his decision depending on voting partial results.

## 4.5 The Anonymity with Secrecy of the Election:
The voter identity should not been known except of the voter himself. So no one can access the voter identity under any condition.

## 4.6 Correctness
The final results of the process of elections must be counted in a correct matter in order to be published.

## 4.7 Verifying Results:
The step of verifying the results comes after finishing the process of tallying and the verification process starts once the results were published. The System must introduce a details of verifying the election results.

## 4.8 Robustness:
The voting system should be able to handle ineligible votes and the votes which cause faults. Some Attackers could participate in casting malicious votes and ballots so the system should be able to recognize these attacks and cancel their effect on the voting process or any server attacks.

## 4.9 The Concern of Coercion:
One of the challenges in a voting system is the possibility to ensure that the user cast his voice without giving his vote to a specific candidate by force even not to let the user to show his vote to anyone else in order not to be able to proof that he or she has voted for specific candidate and get paid for this choice from third party. So the system should be resistant to any coercion [6].

# Chapter 5

# The Cryptographic in Voting Systems

Cryptographic primitives will be described in this chapter in order to get more into cryptographic sphere in voting systems.

## 5.1 Cryptography Public Key

This cryptographic primitive is to manage the voter privacy. The technique works like following:

Each voter has two keys, one is "public key" and the other is "private key".

Every voter uses the "public key", which is in the election's "public key", to encrypt his vote or to encrypt the ballot and this "public key" is published publicly then the voter will use the "private key" to sign the ballot which is already encrypted by the "public key" [9].

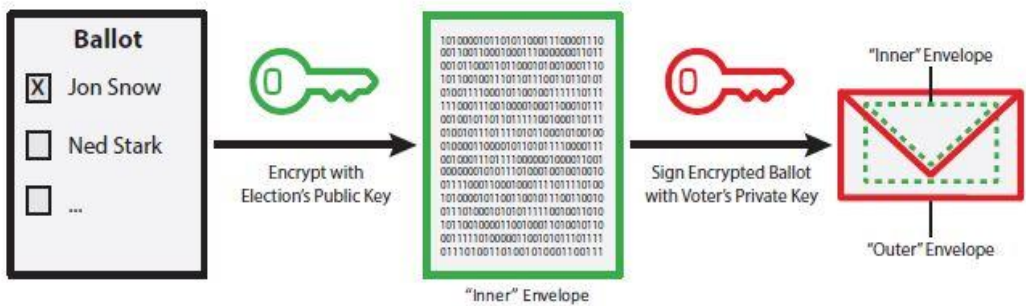

Figure 5.1 singing and encrypting the voting ballot

In order to guarantee the voter anonymity, Mix net technique should be used [10]. One important point while using public key cryptography and RSA, the choosing must be made for safe and secure algorithms to be used in random numbers generator [11] [12].

## 5.2 The Mix Net Property

The voting system can use this property in order to remove some layer of the encryption and then mix and change the order of the votes and send the result to the next node the votes after the votes has been encrypted [13] [14].

To guarantee the voter anonymity there should exist at least one mix net

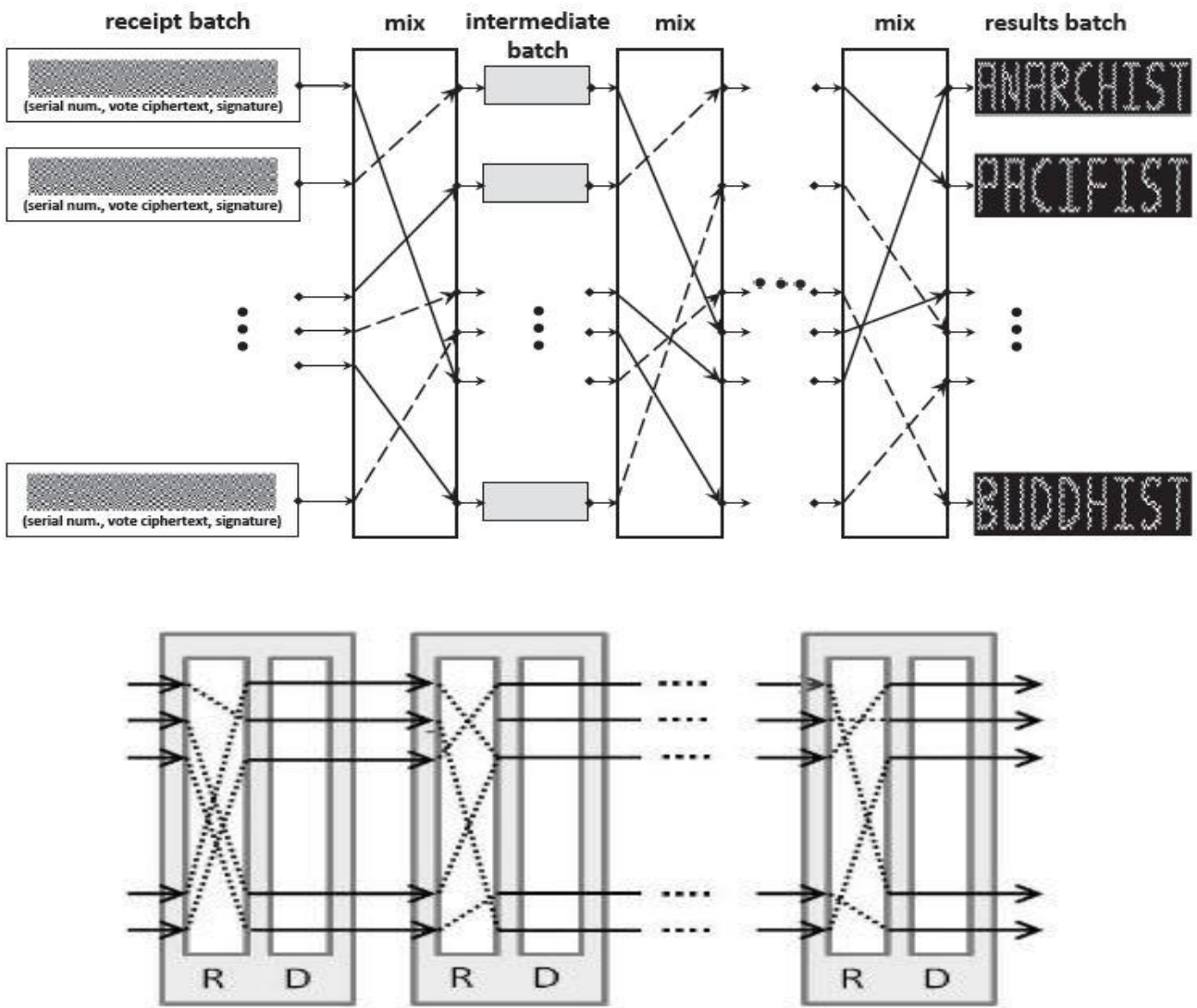

Figure 5.2 The Mix net Property

This property includes one or more mixes which are connected to each other with a specific order (a cascading one) so the outputs of one mix net is the input for the next one to provide voting in anonymous matter [1].

## 5.3 Zero Knowledge Proof

Researchers at MIT in 1980 propose the use of zero knowledge, Goldwasser, Rackoff and Micali [16]. Their research area was about "interactive proof systems" which is explaining how two parties (prover and verifier) can send and receive messages from the prover and verifier in order to make the verifier agree that a specific mathematical statement is totally true.

ZKP must have this properties:

- **Completeness:**

    If the statement happens in honest way "true statement" then it will work as it's expected to and the one who verify will be convinced with this statement by the honest who proved that. A completeness error can be exist because verification can happen with a probability near to one but not totally equal to 1 so the error can be exist. The same scenario can be with the public key encryption during the decryption of the messages [17].

- **ZK:**

    The verifier will not get any knowledge and no information will be gathered except that the truth of the statement and this property represent the actual meaning for the proof zero knowledge [17]. The non-interactive ZKP is used by most of voting systems due to the fact that there is no need for two active parts in the system so the voter is just one part needed to verify different steps in the voting system [19].

- **Soundness:**

If the statement has a false value then the one who prove is not able to convince the one who verify even if there was a cheating from the prover side [18].

The "non-interactive Zero knowledge proofs" are more recommended to be used in the voting systems because in the non-interactive approach the voter is able to process the verification for a lot of steps without the needs of active part in the system and there is no need to take a lot of resources for proofs just the initial one which is used for creating the proofs [19].

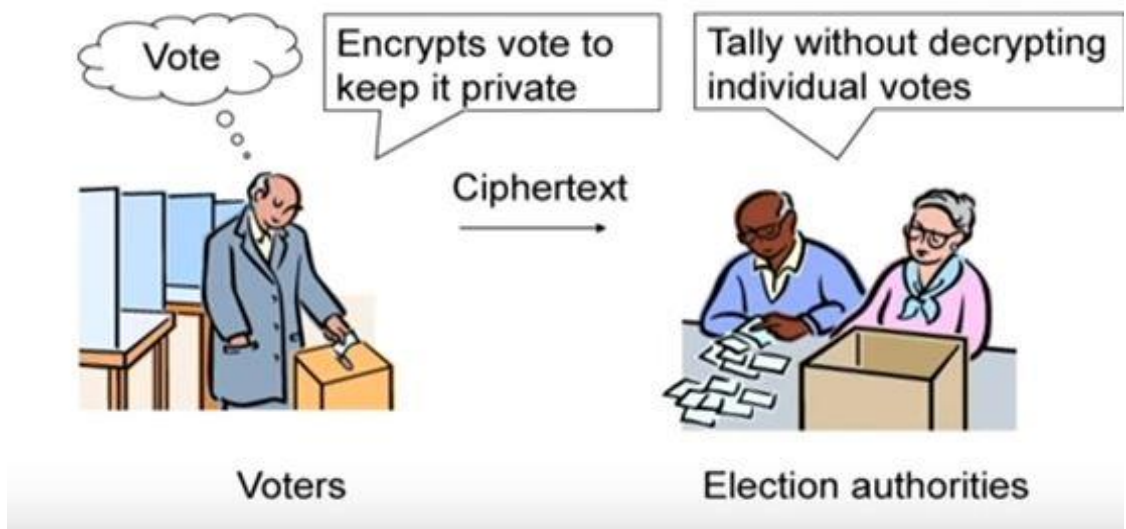

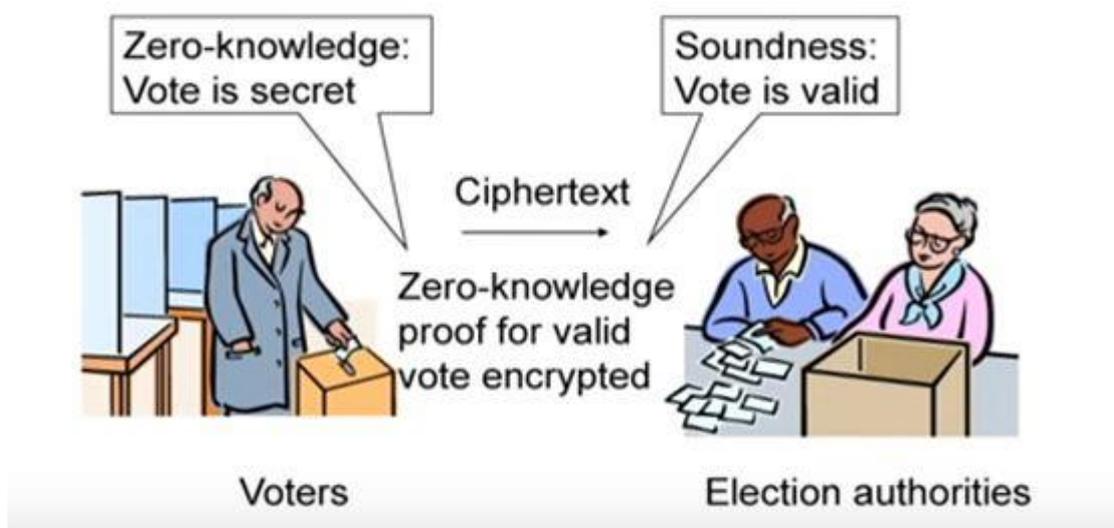

Figure 5.3 NIZKP.

## 5.4 Digital Signatures

A "digital signature" is one of the critical "cryptographic primitives" in order to build blocks. A "digital signature" is a signature in digital form. There are two benefits of a digital signature.

One is the idea of having the distinct signature which can be verified by another party and confirmed that it is a valid signature.

The second is to have the signature to sign different documents or agreements with it and not to be modified by another party except the owner of the signature.

The question is how to create a digital signature with the use of cryptography?

There are three steps to consider:

1) Generate private and public keys, the owner will use the private key as a secret key to create the signature and the public key as a verification key that can be seen by anyone to verify the owner of the signature.

2) Assigning a signature for a particular message that the sender wants it to send in a secret way. A sequence of bits represents the signature here [21].

3) Verifying the signature to be valid or not by using the "public key" of the singer with the message which the signature is on it.

```
(sk, pk) := generateKeys(keysize)
    sk: secret signing key
    pk: public verification key        ⎫
                                       ⎬ can be
sig := sign(sk, message)               ⎭ randomized
                                           algorithms

isValid := verify(pk, message, sig)
```

Figure 4.4 Digital Signature API.

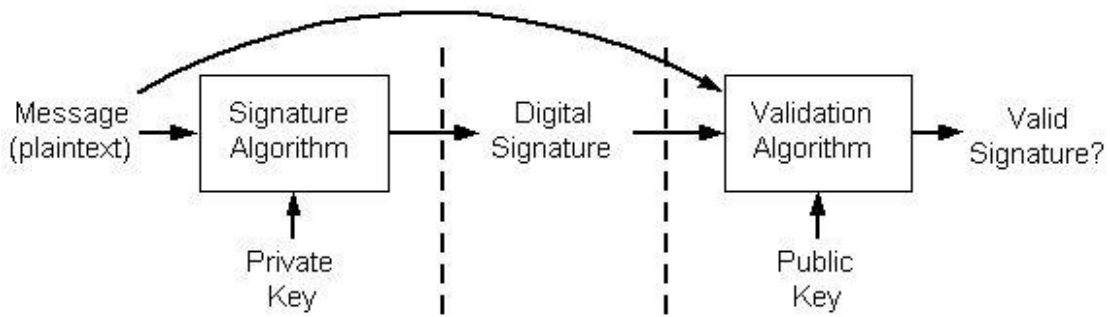

Figure 4.5 Digital Signature mechanism.

Random algorithms can be used in the first two steps but not in the Verification step because it's a deterministic one.

# Chapter 6

# The proposal of a Voting System Architecture

## 6.1 Proposal Aim:

The main goal will be focusing on creating e-voting system with "blockchain technology" to reflect that in the process of making decisions such as:

- Approving a combination of two companies, or new investment.
- Choosing the right directors to the board.
- Approving equity compensation plans.
- Elections of Shareholders.

By using "blockchain technology" for voting issues, every voter will have the ability to verify their own vote, to verify and check that the total votes are accurate, all while remaining as anonymous as they desire. To make the "blockchain technology" more secure, it needs more and more participants.

## 6.2 Waves Platform:

Wave's platform is a blockchain system which is an entirely open source in a decentralized manner to provide full functionalities to transfer, issue and exchange new assets in different cryptocurrencies such as Bitcoin and Ethereum and another cryptocurrency. The system is auditable, decentralized and transparent. There is no need to download the whole blockchain which gives a high accessibility.

Current Waves block maximum size is 100 transactions and a new block can be generated every minute so the speed is up to 100 transactions per minute. The Block speed is one minute and every block can handle around one hundred transaction and also the transactions fees are possible be paid as tokens.

By this specifications, the system will be able to expand from small-scale voting to large-scale voting.

## 6.3 System challenges and requirements:

a) Voters Privacy and how it is possible for every voter to have the ability to verify and check if his/her vote was casted and counted in a correct way or not.

b) Remove any indication of the voter's identity if it's necessary.

c) Unfinished voting results should not be allowed in the system because this can affect the process of voting due to the fact that a voter will vote to the candidate or to choose the answer which has more votes than others.

d) Decreasing threats and ensure that the voter is always voting in safe way without any external environment which can affect the voter decision.

e) Blank votes cannot be used to elect a candidate or make a decision.

f) Using redundant servers in different locations in order to distribute the voting system. So if we have a big number of nodes which are actively participate in the network, any attacker will need a big amount of these malicious nodes to be able to get an impact on blockchain integrity.

g) Choosing algorithms which provide safety in the case of random number generation.

## 6.4 The main elements of a blockchain-based voting system:
### 1. Registration

Each party (which wants to create a poll) and each eligible voter will need their own private wave's wallet. So the voter will create his/her own wallet and the system will verify their eligibility.

### 2. Creating the Issue or the matter to be voted for

Set up the election process details and define the main area of voting. Addresses of each poll and each specific answer will be available publicly.

### 3. Voting transactions Process

The voter can't vote without spending a specific amount of waves, assets or currency (will be decided depending on the poll creation requirements). So the total amount of waves, assets or currency of each answer will be the final results.

### 4. Verifiability

Verification can be done by checking the identification number of the voter, the password and the list of voters.

Each user can see if the vote has arrived in the candidate's wallet. Also, all other transactions can be verified this way to reconstruct the results of the election.

Each candidate specifies a Bitcoin/Wave address. Voters then cast by sending a payment to the selected candidate. Any try to break the voting rules (e.g., one vote per voter) can be noticed by inspecting the blockchain, and the tally is visible by inspecting the candidate's received payments.

Votes should be packed into packages with a defined maximum size and verified with a specific hash. This hash must begin with a certain number of zeros, which depends on the number of participants in the network.

## 6.5 Replace the coin with a vote:

In a PoS "proof-of-stake" system, the holding tokens (in this case waves) will be alternative of hashing power in the mining process.

There are many possible Ways to create a voting system based on blockchain technology.

The simplest one is instead of transferring tokens between accounts, the tokens which transferred in the network can be used to describe individual votes by transferring them into ballots.

The system can have voting right assets and voting token assets for each shareholder. A voter will be able to spend voting tokens to cast their votes on each meeting agenda item if the voter also own the voting right asset.

## 6.6 How to handle double voting:

Before the transactions are added to the blockchain, the inputs of the transactions are checked and it is ensured that these inputs have not been voted before to prevent double voting. The protocol's design defines that the longest chain is the "true" chain. Smaller chains are ignored. Combined with a timestamp and the proof-of-stake, this prevents double voting.

## 6.7 How does the voting system works

The waves voting system will enable the creation of a voting question "Poll" for any valid account with the range of one voting question to one hundred answers. A condition can be added to the system in order to make the user eligible to participate in the voting system if each user have a minimum amount of waves, currency or asset. For each answer, there will be an integer number between specific range values and each answer will have a specific weight depending on different models of the voting process "account model, account model with balance, asset model with balance and the currency model with balance". All of this will be during the creation of the voting question "Poll". After that the result will be by counted as the sum of different weights for all voters multiplied with the integer value for the answers which is chosen when the user cast his/her vote. The votes are saved as attachments and after the ending of elections, the votes will be removed from the blockchain and only remaining is the results.

## 6.8 Voting and creating Poll as Transactions:

- Every "poll transaction or "vote transaction" will need to perform just one operation and then this transaction will be stored on the p2p network in a permanent matter in the block.

- The fees of every "poll transaction" or "vote transaction" are the main and prime technique where the waves are reprocessed back into the p2p network. So each "poll transaction or "vote transaction" will require a 1 wave as minimum fee.

- The transaction cannot be confirmed unless the transaction is totally added into a block which has a valid status.

- There is a parameter called "deadline parameter" which has a specific time in minutes which represent the time when the transaction has been totally submitted into the p2p network.

## 6.9 Transactions Types:

All "poll transaction or "vote transaction" have different parameters :

- A "private key" which represent the voter account
- The specific value of transaction which is "transaction fee".
- A specific time which represent the deadline of making a new voting transaction.
- Optional choice to a referenced transaction.

Waves voting system will be represented as a new transaction type which accept attachment as an input parameter and different parameters depending on the voting area with different processing methods such as"

- The creation of a voting question "poll creation".
- The casting of the vote.

If the user account who is participating in the voting process has enough funds for "the creation of poll " or "vote casting":

    (a) When the new vote/poll transaction is initialized, the every "poll transaction" or "vote transaction" Id will be generated with including the different parameters.

    (b) Using the voter "private key" to sign the transaction.

    (c) Processing the "vote transaction" by putting the encrypted transaction within the p2p network.

    (d) Broadcasting the "poll transaction" or "vote transaction" to all p2p nodes in the network.

    (e) The server is responding with the code of the total results of election, so in case the creation of "voting transaction" was successful then the code will be the "voting transaction ID", otherwise it will be an message which represent the error and fail happens while checking the parameters of the transaction.

## 6.10 Encryption:

As encryption criteria, waves voting system will encrypt the transactions which are included in the voting procedure by using "Elliptic-Curve Korean Certificate based Digital Signature Algorithm (EC-KCDSA)".

## 6.11 Voter use case diagram:

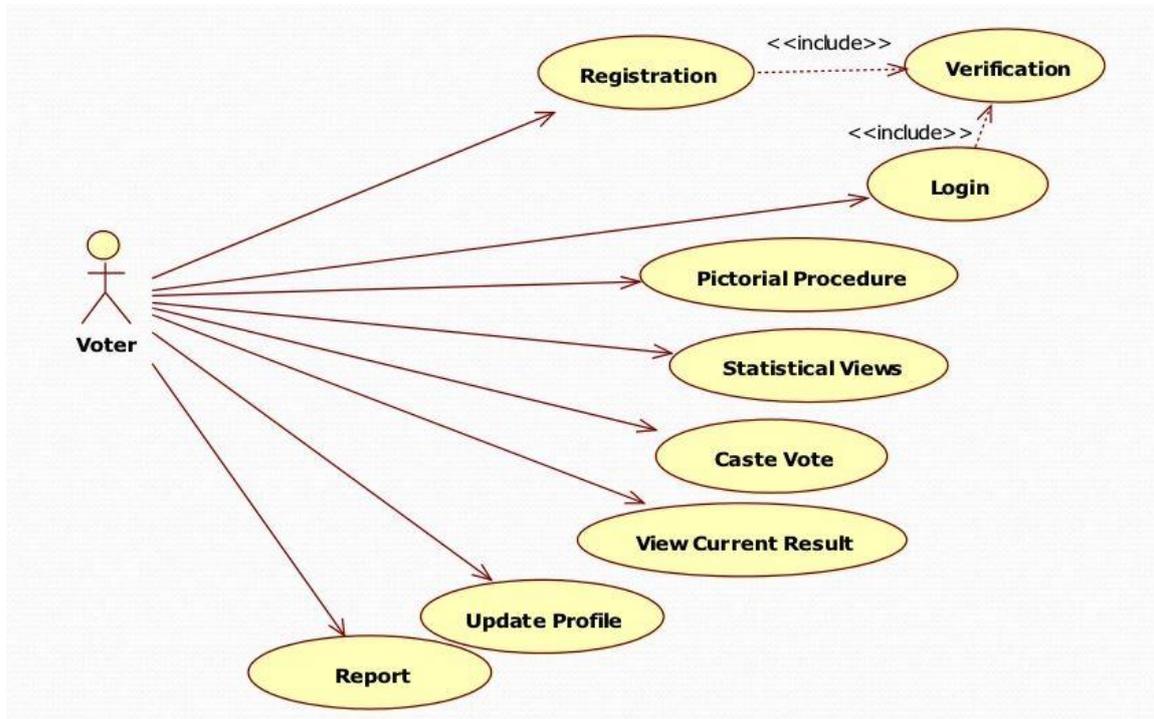

## 6.12 Sequence diagram for the voting system:

### Login Process:

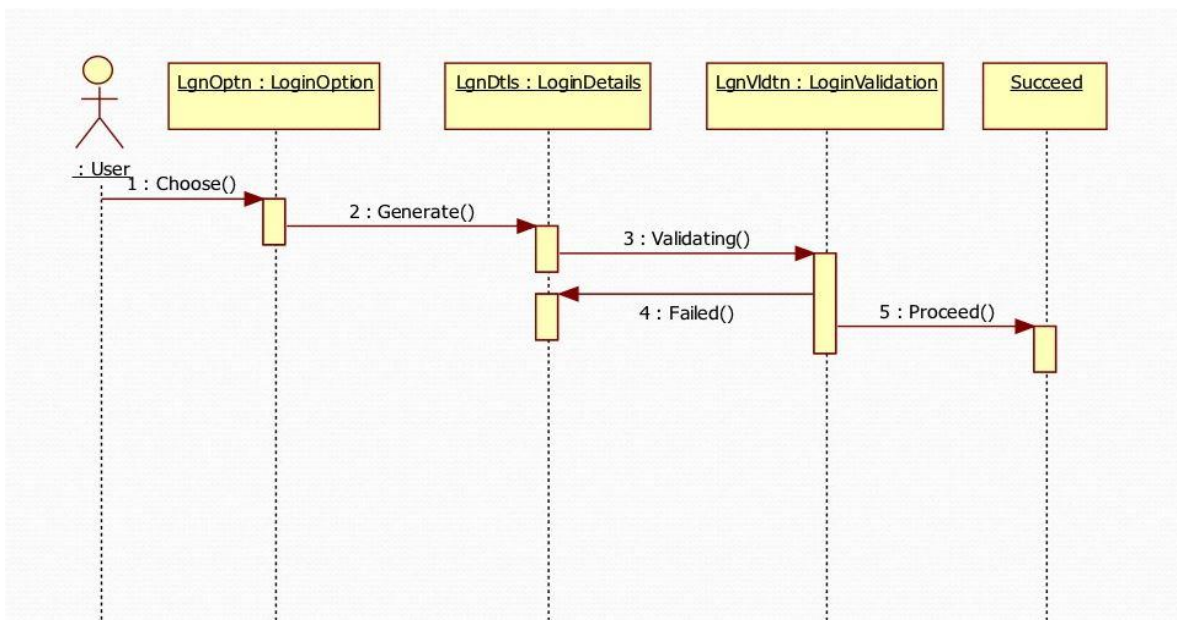

## Voting process:

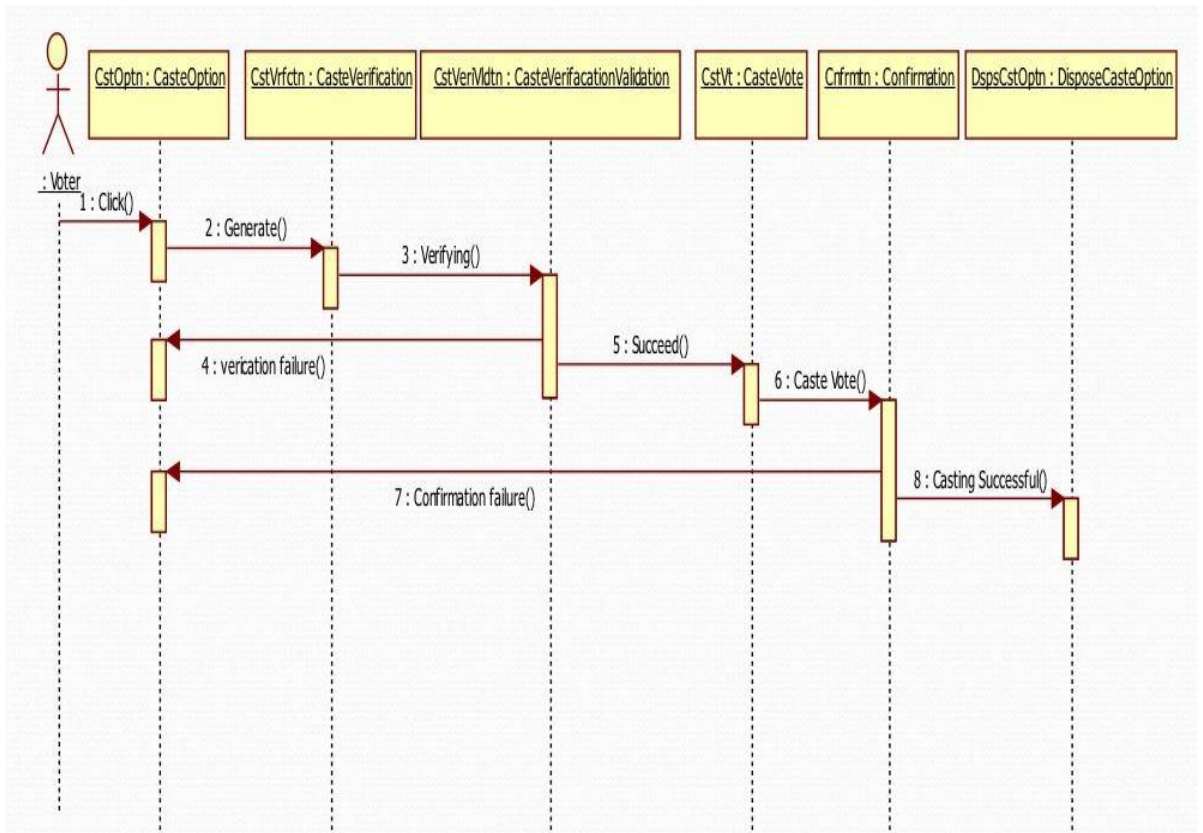

## Results process:

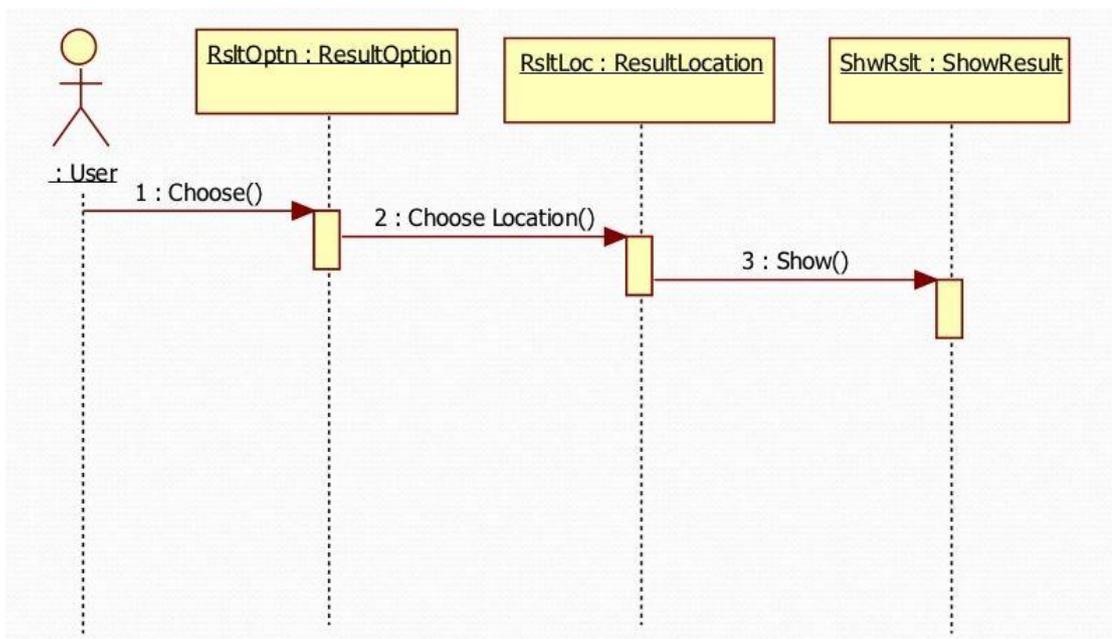

# Chapter 7

## Conclusion and Future Work

In this research, various electronic voting systems were studied and a new architecture was proposed through the use of proof of stake protocol which gives the possibility to have secure system without depending on massive computational power as in proof of work protocol which is used by Ethereum blockchain to find the hashes. The process of designing the system is handling the security issues that are needed in real voting systems.

The proposal used waves platform as a blockchain system to bring the ideas of an electronic voting system which use PoS to the real world.

This will consider the use of smartphones and small devices to take part in the election process in the peer to peer network to ensure the complete integrity of the whole blockchain. This can be reached in an easy way by allowing the smartphones to be online in order to be a full node in the voting peer to peer network which is a blockchain system.

As future work, the proposed architecture will be implemented on wave lite client to provide a real voting product which at the beginning will focus on shareholders elections and then go further for national election over waves blockchain. To create this type of large voting system, a dedicated blockchain will be responsible only for handling the voting process with big block size to handle a lot of transaction on the chain and centralized maintenance.

# Appendix:

```scala
package scorex.transaction.Data

import com.google.common.base.Charsets
import com.google.common.primitives.{Bytes, Longs}
import play.api.libs.json.{JsObject, Json}
import scorex.account.{PrivateKeyAccount, PublicKeyAccount}
import scorex.crypto.EllipticCurveImpl
import scorex.crypto.EllipticCurveImpl.SignatureLength
import scorex.crypto.encode.Base58
import scorex.serialization.BytesSerializable
import scorex.transaction._
import scorex.transaction.TransactionParser._

import scala.util.{Failure, Success, Try}
import scala.util.Try

/**
  * Created by DN on 30/05/2017.
  */
sealed trait DataTransaction extends SignedTransaction
{
  def data: Array[Byte]
  def fee: Long
  def dataLength: Int
}

object DataTransaction
{
  private case class  DataTransactionImpl(sender: PublicKeyAccount,
                                          data: Array[Byte],
                                          dataLength: Long,
                                          fee: Long,
                                          timestamp:Long,
                                          signature: Array[Byte])
    extends DataTransaction
  {
    override val transactionType: TransactionType.Value = TransactionType.DataTransaction
    override val assetFee: (Option[AssetId], Long)   = (None, fee)

    lazy val toSign: Array[Byte] =
Bytes.concat(Array(transactionType.id.toByte),
                                             sender.publicKey,
BytesSerializable.arrayWithSize(data),
                                             Array(dataLength.toByte),
                                             Longs.toByteArray(fee),
Longs.toByteArray(timestamp))

    override lazy val json: JsObject = jsonBase() ++ Json.obj(
      "data" -> Base58.encode(data)
    )

    override lazy val bytes: Array[Byte] =
Bytes.concat(Array(transactionType.id.toByte), signature, toSign)

  }

  val MaxDataSize = 140
  def parseTail(bytes: Array[Byte]): Try[DataTransaction] = Try {
    val signature = bytes.slice(0, SignatureLength)
    val txId      = bytes(SignatureLength)
    require(txId == TransactionType.DataTransaction.id.toByte, s"Signed tx
```

```scala
id is not match")
    val sender                           = PublicKeyAccount(bytes.slice(SignatureLength + 1, SignatureLength + KeyLength + 1))
    val (dataLength, dataStart)          = Array()
    val data                             = b
    val fee                              = Longs.fromByteArray(bytes.slice(dataStart + 10, dataStart + 18))
    val timestamp                        = Longs.fromByteArray(bytes.slice(data + 18, data + 26))
    DataTransaction.create(sender,dataLength,data, fee, timestamp, signature)
      .fold(left => Failure(new Exception(left.toString)), right => Success(right))
  }.flatten

  private def createUnverified(sender: PublicKeyAccount,
                               data: Array[Byte],
                               dataLength: Long,
                               fee: Long,
                               timestamp:Long,
                               signature: Option[Array[Byte]] = None) =

    if (dataLength > MaxDataSize) {
      Left(ValidationError.TooBigArray)
    }  else if (fee <= 0) {
      Left(ValidationError.InsufficientFee)
    } else {
      Right(DataTransactionImpl(sender, data, dataLength, fee, timestamp, signature.orNull))
    }

  def create(sender: PublicKeyAccount,
             data: Array[Byte],
             dataLength: Long,
             fee: Long,
             timestamp:Long,
             signature: Array[Byte]): Either[ValidationError, DataTransaction] =
    createUnverified(sender, data, dataLength , fee, timestamp, Some(signature))
      .right.flatMap(SignedTransaction.verify)

  def create(sender: PrivateKeyAccount,
             data: Array[Byte],
             dataLength: Long,
             fee: Long,
             timestamp: Long): Either[ValidationError, DataTransaction] =
    createUnverified(sender, data, dataLength , fee, timestamp).right.map { unverified =>
      unverified.copy(signature = EllipticCurveImpl.sign(sender, unverified.toSign))
    }

}
```

```scala
package scorex.transaction.assets

import com.google.common.base.Charsets
import com.google.common.primitives.{Bytes, Longs}
import play.api.libs.json.{JsObject, Json}
import scorex.account.{Account, PrivateKeyAccount, PublicKeyAccount}
import scorex.crypto.EllipticCurveImpl
import scorex.crypto.encode.Base58
import scorex.serialization.{BytesSerializable, Deser}
import scorex.transaction.TransactionParser._
import scorex.transaction.ValidationError
import scorex.transaction._

import scala.util.{Failure, Success, Try}

sealed trait IssueTransaction extends AssetIssuance {
  def name: Array[Byte]
  def description: Array[Byte]
  def decimals: Byte
  def fee: Long
}

object IssueTransaction {

  private case class IssueTransactionImpl(sender: PublicKeyAccount,
                                          name: Array[Byte],
                                          description: Array[Byte],
                                          quantity: Long,
                                          decimals: Byte,
                                          reissuable: Boolean,
                                          fee: Long,
                                          timestamp: Long,
                                          signature: Array[Byte])
    extends IssueTransaction {

    override val assetFee: (Option[AssetId], Long)     = (None, fee)
    override val transactionType: TransactionType.Value =
TransactionType.IssueTransaction

    override lazy val assetId = id

    lazy val toSign: Array[Byte] =
Bytes.concat(Array(transactionType.id.toByte),
                                                       sender.publicKey,

BytesSerializable.arrayWithSize(name),

BytesSerializable.arrayWithSize(description),
                                                       Longs.toByteArray(quantity),
                                                       Array(decimals),
                                                       if (reissuable) Array(1:
Byte) else Array(0: Byte),
                                                       Longs.toByteArray(fee),

Longs.toByteArray(timestamp))

    override lazy val json: JsObject = jsonBase() ++ Json.obj(
        "assetId"     -> Base58.encode(assetId),
        "name"        -> new String(name, Charsets.UTF_8),
        "description" -> new String(description, Charsets.UTF_8),
        "quantity"    -> quantity,
        "decimals"    -> decimals,
        "reissuable"  -> reissuable
      )

    override lazy val bytes: Array[Byte] =
Bytes.concat(Array(transactionType.id.toByte), signature, toSign)
```

```scala
    }

    val MaxDescriptionLength = 1000
    val MaxAssetNameLength   = 16
    val MinAssetNameLength   = 4
    val MaxDecimals          = 8

    def parseTail(bytes: Array[Byte]): Try[IssueTransaction] = Try {
      val signature = bytes.slice(0, SignatureLength)
      val txId      = bytes(SignatureLength)
      require(txId == TransactionType.IssueTransaction.id.toByte, s"Signed tx id is not match")
      val sender                        = PublicKeyAccount(bytes.slice(SignatureLength + 1, SignatureLength + KeyLength + 1))
      val (assetName, descriptionStart) = Deser.parseArraySize(bytes, SignatureLength + KeyLength + 1)
      val (description, quantityStart)  = Deser.parseArraySize(bytes, descriptionStart)
      val quantity                      = Longs.fromByteArray(bytes.slice(quantityStart, quantityStart + 8))
      val decimals                      = bytes.slice(quantityStart + 8, quantityStart + 9).head
      val reissuable                    = bytes.slice(quantityStart + 9, quantityStart + 10).head == (1: Byte)
      val fee                           = Longs.fromByteArray(bytes.slice(quantityStart + 10, quantityStart + 18))
      val timestamp                     = Longs.fromByteArray(bytes.slice(quantityStart + 18, quantityStart + 26))
      IssueTransaction.create(sender, assetName, description, quantity, decimals, reissuable, fee, timestamp, signature)
        .fold(left => Failure(new Exception(left.toString)), right => Success(right))
    }.flatten

    private def createUnverified(sender: PublicKeyAccount,
                                 name: Array[Byte],
                                 description: Array[Byte],
                                 quantity: Long,
                                 decimals: Byte,
                                 reissuable: Boolean,
                                 fee: Long,
                                 timestamp: Long,
                                 signature: Option[Array[Byte]] = None) =
      if (quantity <= 0) {
        Left(ValidationError.NegativeAmount)
      } else if (description.length > MaxDescriptionLength) {
        Left(ValidationError.TooBigArray)
      } else if (name.length < MinAssetNameLength || name.length > MaxAssetNameLength) {
        Left(ValidationError.InvalidName)
      } else if (decimals < 0 || decimals > MaxDecimals) {
        Left(ValidationError.TooBigArray)
      } else if (fee <= 0) {
        Left(ValidationError.InsufficientFee)
      } else {
        Right(IssueTransactionImpl(sender, name, description, quantity, decimals, reissuable, fee, timestamp, signature.orNull))
      }

    def create(sender: PublicKeyAccount,
               name: Array[Byte],
               description: Array[Byte],
               quantity: Long,
               decimals: Byte,
               reissuable: Boolean,
               fee: Long,
               timestamp: Long,
```

```scala
                      signature: Array[Byte]): Either[ValidationError,
IssueTransaction] =
    createUnverified(sender, name, description, quantity, decimals,
reissuable, fee, timestamp, Some(signature))
      .right.flatMap(SignedTransaction.verify)

  def create(sender: PrivateKeyAccount,
             name: Array[Byte],
             description: Array[Byte],
             quantity: Long,
             decimals: Byte,
             reissuable: Boolean,
             fee: Long,
             timestamp: Long): Either[ValidationError, IssueTransaction] =
    createUnverified(sender, name, description, quantity, decimals,
reissuable, fee, timestamp).right.map { unverified =>
      unverified.copy(signature = EllipticCurveImpl.sign(sender,
unverified.toSign))
    }
}
```

```scala
package scorex.transaction.assets

import scala.util.{Failure, Success, Try}
import com.google.common.primitives.{Bytes, Longs}
import com.wavesplatform.utils.base58Length
import play.api.libs.json.{JsObject, Json}
import scorex.account.{Account, AccountOrAlias, PrivateKeyAccount,
PublicKeyAccount}
import scorex.crypto.EllipticCurveImpl
import scorex.crypto.encode.Base58
import scorex.serialization.{BytesSerializable, Deser}
import scorex.transaction.TransactionParser._
import scorex.transaction.{ValidationError, _}

sealed trait TransferTransaction extends SignedTransaction {
  def assetId: Option[AssetId]

  def recipient: AccountOrAlias

  def amount: Long

  def feeAssetId: Option[AssetId]

  def fee: Long

  def attachment: Array[Byte]
}

object TransferTransaction {

  val MaxAttachmentSize = 140
  val MaxAttachmentStringSize = base58Length(MaxAttachmentSize)

  private case class TransferTransactionImpl(assetId: Option[AssetId],
                                             sender: PublicKeyAccount,
                                             recipient: AccountOrAlias,
```

```scala
                                                        amount: Long,
                                                        timestamp: Long,
                                                        feeAssetId: Option[AssetId],
                                                        fee: Long,
                                                        attachment: Array[Byte],
                                                        signature: Array[Byte])
    extends TransferTransaction {
    override val transactionType: TransactionType.Value = TransactionType.TransferTransaction

    override val assetFee: (Option[AssetId], Long) = (feeAssetId, fee)

    lazy val toSign: Array[Byte] = {
      val timestampBytes = Longs.toByteArray(timestamp)
      val assetIdBytes = assetId.map(a => (1: Byte) +: a).getOrElse(Array(0: Byte))
      val amountBytes = Longs.toByteArray(amount)
      val feeAssetIdBytes = feeAssetId.map(a => (1: Byte) +: a).getOrElse(Array(0: Byte))
      val feeBytes = Longs.toByteArray(fee)

      Bytes.concat(Array(transactionType.id.toByte),
        sender.publicKey,
        assetIdBytes,
        feeAssetIdBytes,
        timestampBytes,
        amountBytes,
        feeBytes,
        recipient.bytes,
        BytesSerializable.arrayWithSize(attachment))
    }

    override lazy val json: JsObject = jsonBase() ++ Json.obj(
      "recipient" -> recipient.stringRepr,
      "assetId" -> assetId.map(Base58.encode),
      "amount" -> amount,
      "feeAsset" -> feeAssetId.map(Base58.encode),
      "attachment" -> Base58.encode(attachment)
    )

    override lazy val bytes: Array[Byte] = Bytes.concat(Array(transactionType.id.toByte), signature, toSign)

  }

  def parseTail(bytes: Array[Byte]): Try[TransferTransaction] = Try {
    import EllipticCurveImpl._

    val signature = bytes.slice(0, SignatureLength)
    val txId = bytes(SignatureLength)
    require(txId == TransactionType.TransferTransaction.id.toByte, s"Signed tx id is not match")
    val sender = PublicKeyAccount(bytes.slice(SignatureLength + 1, SignatureLength + KeyLength + 1))
    val (assetIdOpt, s0) = Deser.parseOption(bytes, SignatureLength + KeyLength + 1, AssetIdLength)
    val (feeAssetIdOpt, s1) = Deser.parseOption(bytes, s0, AssetIdLength)
    val timestamp = Longs.fromByteArray(bytes.slice(s1, s1 + 8))
    val amount = Longs.fromByteArray(bytes.slice(s1 + 8, s1 + 16))
    val feeAmount = Longs.fromByteArray(bytes.slice(s1 + 16, s1 + 24))

    (for {
      recRes <- AccountOrAlias.fromBytes(bytes, s1 + 24)
      (recipient, recipientEnd) = recRes
      (attachment, _) = Deser.parseArraySize(bytes, recipientEnd)
      tt <- TransferTransaction.create(assetIdOpt, sender, recipient, amount, timestamp, feeAssetIdOpt, feeAmount, attachment, signature)
```

```scala
    } yield tt).fold(left => Failure(new Exception(left.toString)), right =>
Success(right))
  }.flatten

  private def createUnverified(assetId: Option[AssetId],
                               sender: PublicKeyAccount,
                               recipient: AccountOrAlias,
                               amount: Long,
                               timestamp: Long,
                               feeAssetId: Option[AssetId],
                               feeAmount: Long,
                               attachment: Array[Byte],
                               signature: Option[Array[Byte]] = None) = {
    if (attachment.length > TransferTransaction.MaxAttachmentSize) {
      Left(ValidationError.TooBigArray)
    } else if (amount <= 0) {
      Left(ValidationError.NegativeAmount) //CHECK IF AMOUNT IS POSITIVE
    } else if (Try(Math.addExact(amount, feeAmount)).isFailure) {
      Left(ValidationError.OverflowError) // CHECK THAT fee+amount won't overflow Long
    } else if (feeAmount <= 0) {
      Left(ValidationError.InsufficientFee)
    } else {
      Right(TransferTransactionImpl(assetId, sender, recipient, amount, timestamp, feeAssetId, feeAmount, attachment, signature.orNull))
    }
  }

  def create(assetId: Option[AssetId],
             sender: PublicKeyAccount,
             recipient: AccountOrAlias,
             amount: Long,
             timestamp: Long,
             feeAssetId: Option[AssetId],
             feeAmount: Long,
             attachment: Array[Byte],
             signature: Array[Byte]): Either[ValidationError, TransferTransaction] = {
    createUnverified(assetId, sender, recipient, amount, timestamp, feeAssetId, feeAmount, attachment, Some(signature))
      .right.flatMap(SignedTransaction.verify)
  }

  def create(assetId: Option[AssetId],
             sender: PrivateKeyAccount,
             recipient: AccountOrAlias,
             amount: Long,
             timestamp: Long,
             feeAssetId: Option[AssetId],
             feeAmount: Long,
             attachment: Array[Byte]): Either[ValidationError, TransferTransaction] = {
    createUnverified(assetId, sender, recipient, amount, timestamp, feeAssetId, feeAmount, attachment).right.map { unsigned =>
      unsigned.copy(signature = EllipticCurveImpl.sign(sender, unsigned.toSign))
    }
  }
}
```